\documentclass[fleqn,usenatbib]{mnras}
\usepackage{newtxtext,newtxmath}

\usepackage[T1]{fontenc}
\usepackage{ae,aecompl}

\usepackage{graphicx}	
\usepackage{amsmath}	
\usepackage{amssymb}	

\newcommand{\dg}{{DG92-311}}

\title[Varibility in A851]{The nature of the variable millimetre--selected AGN in the Brightest Cluster Galaxy of {Abell~851}}

\author[Cheale et al.]{R.~A.~Cheale,$^{1}$\thanks{E-mail: r.cheale@herts.ac.uk (RAC)}
J.~E.~Geach,$^{1}$
A.~C.~Edge,$^{2}$ 
Y.~C.~Perrott,$^{3}$
T.~Cantwell$^{4}$
\\
$^{1}$Center for Astrophysics Research, School of Physics, Astronomy \& Mathematics, University of Hertfordshire, Hatfield, AL10 9AB\\
$^{2}$Centre for Extragalactic Astronomy, Department of Physics, Durham University, South Road,  Durham, DH1 3LE\\
$^{3}$Astrophysics Group, Cavendish Laboratory, 19 J. J. Thomson Avenue, Cambridge CB3 0HE\\
$^{4}$Jodrell Bank Centre for Astrophysics, Alan Turing Building, School of Physics and Astronomy, University of Manchester, M13 9PL
}

\date{Accepted XXX. Received YYY; in original form ZZZ}

\pubyear{2018}

\begin{document}
\label{firstpage}
\pagerange{\pageref{firstpage}--\pageref{lastpage}}
\maketitle

\begin{abstract}
We present the detection of a bright 3\,mm continuum source in the Brightest Cluster Galaxy (BCG) in Abell~0851 ($z=0.411$) with the NOrthern Extended Millimeter Array (NOEMA). When this detection is compared to other multi-frequency observations across $21$cm--$100\micron$, including new Arcminute Microkelvin Imager $15$\,GHz observations, we find evidence for a relatively flat, variable core source associated with the BCG. The radio power and amplitude of variability observed in this galaxy is consistent with the cores in lower redshift BCGs in X-ray--selected clusters, and the flat mm--cm spectrum is suggestive of the BCG being a low luminosity AGN archetype. The discovery of this system could provide a basis for a long-term study of the role of low luminosity radio mode `regulatory' feedback in massive clusters. 
\end{abstract}

\begin{keywords}
galaxies: clusters: individual: Abell 851 -- galaxies:clusters:individual (A851, Cl 0939+4713) -- galaxies:evolution -- galaxies: elliptical and lenticular, cD -- techniques: interferometric
\end{keywords}



\section{Introduction}

The discovery that every massive galaxy contains a super-massive black hole (SMBH), and that the masses of the stellar bulge and SMBH are correlated \citep{magorrian_demography_1998,silk_quasars_1998} demonstrates that the growth of the central black hole and its host galaxy are inexorably linked. Black hole accretion releases large amounts of feedback energy and momentum into the interstellar medium (and beyond) via collimated jets and fast winds driven from the hot accretion disc, and is thought to be a driving feature in the regulation of stellar mass growth \citep{bower_breaking_2006, croton_many_2006}. Active Galactic Nucleus (AGN) feedback is now an established feature of galaxy formation models that are required to correctly reproduce the key observable features of the local galaxy population \citep[e.g.,][]{sijacki_unified_2007,booth_cosmological_2009,fabian_observational_2012,ishibashi_active_2012}

We observe a wide range of nuclear activity in galaxies; from low luminosity or quiescent systems such as Sgr A$^\star$ at the center of the Milky Way, up to powerful radio galaxies and quasars where AGN feedback can expel large fractions of the gas reservoir and pump energy into the circumgalactic and intergalactic medium \citep{gaspari_agn_2011,dubois_blowing_2013,schaye_eagle_2015}. This substantial energy input into the local environment is necessary for regulating stellar mass growth on galaxy scales by stifling the cooling of intracluster gas\citep[see; ][]{fabian_observational_2012}, but this must be sustained for many Gyr to maintain the suppression of stellar mass growth in the host galaxy \citep{dunn_investigating_2008}. It is likely that a self-regulating process affects the growth of the central black hole, as exhibited in numerical simulations \citep[e.g.,][]{springel_modeling_2005} and galaxy formation theories \citep{silk_quasars_1998,benson_galaxy_2010}.

Variability is associated with all AGN from high luminosity quasars to Seyfert galaxies \citep{mchardy_active_2006}. Low-luminosity AGN (LLAGN), like Sgr A$^\star$, M81 \citep{sakamoto_millimetric_2001, schodel_millimetre_2007}, Centaurus A \citep{israel_millimeter-wave_2008} and, NGC7469 \citep{baldi_milimetre-band_2015} have relatively low Eddington rates ($L \approx 0.1\,L_\textrm{Edd}$) and often exhibit bright inverted/flat cm-mm spectra likely originating from a compact core \citep[e.g.][]{behar_mm-wave_2018}. Recent advances in mm-interferometry in resolution and sensitivity make the detection of such LLAGN more probable. Indeed, work by \citealp{doi_millimeter_2011} supports the view that many large passive galaxies have compact millimetre cores with significant variable radio activity at their cores, but few systematic searches for such LLAGN have yet been made, and in general the detection and monitoring of AGN variability requires multiple observations over weeks to decades. 

Here we present evidence of a variable AGN in the well known cluster {Abell~851} (also known as CL\,0939+4713), a rich\citep{seitz_mass_1996} cluster ($M\sim10^{14}M_\odot$at $z=0.411$ containing several hundred spectroscopically classified members \citep{dressler_spectroscopy_1992}. The galaxy in question is a possible Sa/S0 transition object close to the cluster centre, catalogued by Dressler \& Gunn (1992) as object 311 and hence referred to in this paper as \dg{}. \dg{} is optically classified as an early-type disc (Sa/S0) with post-starburst spectral features; namely weak nebular emission but relatively strong Balmer absorption \citep[(k+a)][]{belloni_detectability_1995,dressler_spectroscopic_1999,oemler_jr_abell_2009}. In this Letter we present the observational evidence across the radio-far-infrared spectral energy distribution of \dg{}, including new 3\,mm IRAM Plateau de Bure interferometer and new 1.9\,cm Arcminute Microkelvin Imager observations. Throughout we assume a $\Lambda$CDM cosmology with $H_{0} = 67.3\,$km s$^{-1}\,$Mpc$^{-1}$, $\Omega_{\textrm{M}} = 0.315$, and $\Omega_{\Lambda} = 0.685$ \citep{planck_collaboration_planck_2016}.

\section{Data Collection and Analysis} \label{sec:data}

\begin{figure}
\centering
\includegraphics[width=\columnwidth]{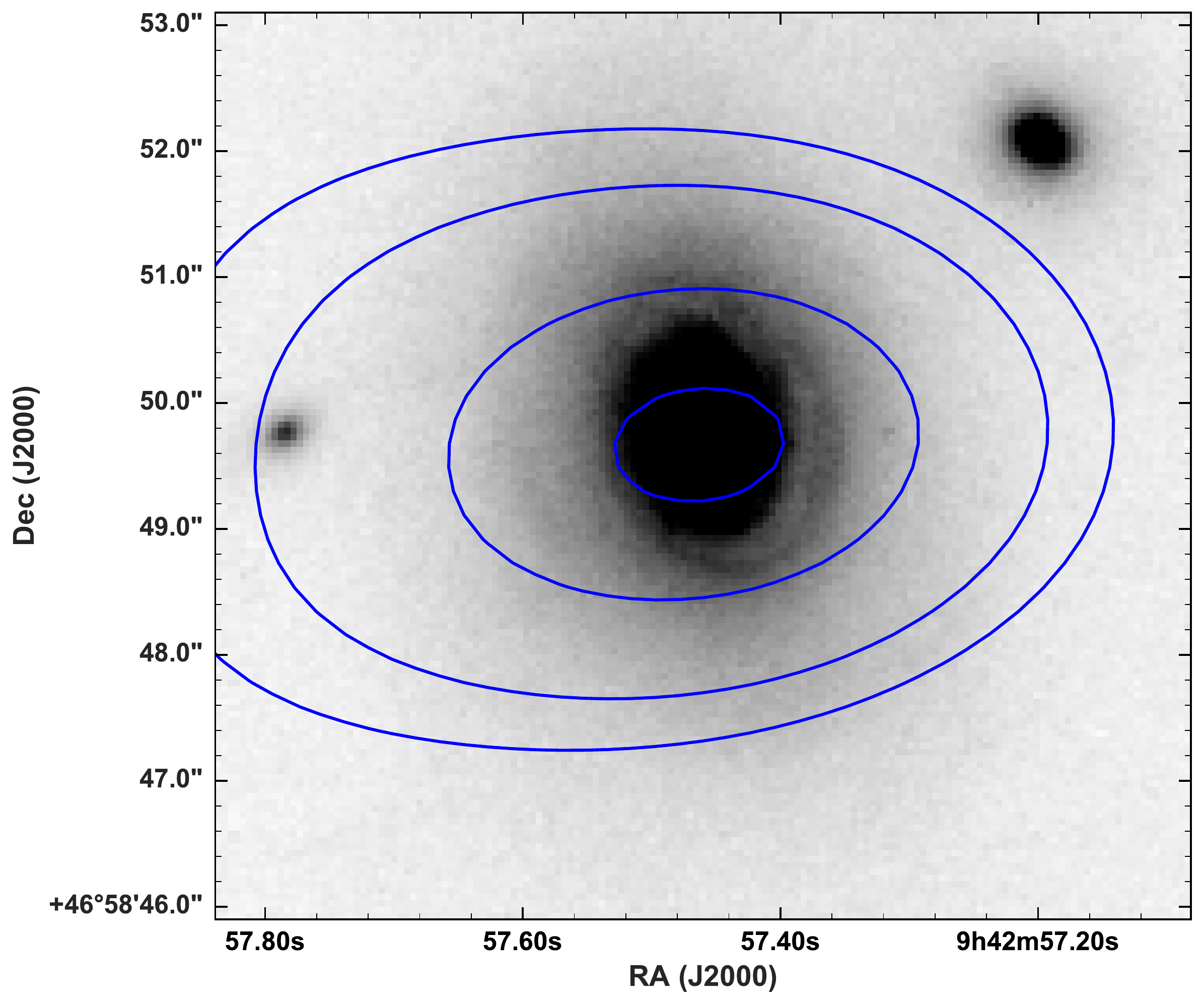}
\caption{The archival {\it HST}/ACS F814W filter optical imaging. {\dg } is located $\alpha;\,09^{\textrm{h}}\,42^{\textrm{m}}\,57.50^{\textrm{s}}$, $\delta;\,46^{\textrm{d}}\,48^{\textrm{m}}\,50.00^{\textrm{s}}$. The primary beam corrected significance contours are overlayed, initiating at $1.05\,$mJy ($3\sigma$) and increasing in $1\sigma$ intervals. We find a corrected (see Section~\ref{subsec:iram}) peak flux density of S$_{3.6\textrm{mm}}1.59\,\pm\,0.15\,$mJy.}
\label{fig:pbCor}
\end{figure}

We make use of a number of archival observations of \dg{} including {\it WISE} $3.4-22\micron$, \textit{Hershel} PACS and SPIRE 100--500$\micron$, James Clerk Maxwell Telescope (JCMT) SCUBA $850\micron$ (project M00BH05), BIMA 1.05\,cm, AMI 1.9\,cm and, VLA/FIRST $6.2-21$\,cm band archives. The {\it HST}/ACS F814W filter optical imaging is shown in Figure~\ref{fig:pbCor} (data retrieved from the MAST archive HST, project 10418). We report the various flux density measurements of \dg{} in Table~\ref{tab:observations}. 

\subsection{Archival Data}

\subsubsection{JCMT/SCUBA}

Observations of {\dg } by SCUBA  at $850\,\micron$ were conducted on the 27th November 2000. The target was observed in the `jiggle' map mode used for observing sources smaller than the array \citep{jenness_scuba_2000}. The target is formally undetected in the archival map, and so we determine an upper limit by sampling a large number of random pixels around the source position and fit a Gaussian to the resulting pixel distribution. We take the standard deviation of the Gaussian as a measure the 1$\sigma$ noise, and determine a 3$\sigma$ upper limit of S$_\textrm{$850\micron$}\,<\,4.6$\,mJy for \dg{}. 

\subsubsection{Herschel Lensing Survey}

The \textit{Herschel} Lensing Survey \citep[HLS][]{egami_herschel_2010} observed Abell\,851 with the PACS ($100$ and $160\,\mu$m) and SPIRE ($250$, $350$ and $500\,\mu$m) instruments. These far-IR/sub-mm bands are useful to constrain the peak of the thermal dust emission. \cite{rawle_relation_2012} analyse a number of BCGs from the HLS including A851. Photometry in the SPIRE bands was reduced by using the \texttt{IRAF} package \texttt{ALLSTAR} \citep{tody_iraf_1993} by fitting the Point Spread Function to source locations. In the PACS bands fluxes were measured by using aperture photometry with the use of {\texttt{SEXTRACTOR}} \citep{bertin_sextractor:_1996}; methods are described in detail in \citeauthor{rawle_relation_2012}. \dg{} was detected in all but the $500\,\micron$ band, and we report the measurements in Table~\ref{tab:observations}.

\subsection{NOEMA observations}\label{subsec:iram}
In project S14BV (PI: Geach), we observed \dg{} as part of a larger 3\,mm mosaic of Abell\,851 to search for co(2--1) emission associated with cluster members (e.g. Geach et al.\ 2009). Abell\,851 was observed in configuration D (baseline separations up to 150\,m) for maximum sensitivity.  We adopted a similar set up to Geach et al.\ (2009), where the 3\,mm receiver was set to the frequency of the redshifted CO(1–0) line at the redshift of the cluster, and the correlator was set up with 2.5-MHz spacing ($2\times64$ channels, 320-MHz bandwidth). The data were reduced using the standard Grenoble Image and Line Data Analysis Software (\textsc{GILDAS}\footnote{http://www.iram.fr/IRAMFR/GILDAS}) and converted to a \textsc{UVFITS} data table for imaging in the \textsc{CASA} environment \citep{mcmullin_casa_2007}. The 3\,mm continuum detection became obvious in a channel-by-channel inspection of the data cube. We note the source lies close to the edge of the 50$''$ primary beam, and we apply an appropriate primary beam correction to measure a flux density $\textrm{S}_\textrm{$3.6\,$mm}\,=\,2.6\,\pm\,0.4\,\textrm{mJy}$. The 3\,mm contours are overlaid on {\it HST}/ACS optical imaging in Figure~\ref{fig:pbCor}. 

\subsection{The Arcminute Microkelvin Imager}

The Arcminute Microkelvin Imager \citep[AMI;][]{zwart_arcminute_2008,hickish_digital_2018} is a dual aperture-synthesis array that operates between $13.9-18.2$GHz with 2048 channels. The principle use of the AMI detector is for imaging the Sunyaev-Zel'dovich effect by observing galaxy clusters. However, we make use of the instrument for its favourable bandwidth for observations pointed at A851. {\dg } was observed by AMI on the 4$^{\textrm{th}}$ October 2017 with an integration time of $7200$\,s, we make use of AMI-LA with angular resolution of $30''$ using seven of the eight 12.8\,m diameter dishes. 

The AMI-LA data were calibrated and imaged in CASA. Primary calibration was performed using a nearby observation of 3C\,286, using the \citealp{perley_accurate_2013} flux density scale along with a correction for the fact that AMI measures I+Q, using the polarization fraction and angle fits from \citealp{perley_integrated_2013}; this is an $\approx\,4.5\%$ correction for 3C\,286 over the AMI band. The primary calibration observation supplied an instrumental bandpass in both phase and amplitude. This was applied to the target data, as well as a correction for atmospheric amplitude variations produced by the `rain gauge', which is a noise injection system used to measure the atmospheric noise contribution \citep[see;][]{zwart_arcminute_2008}. The nearby bright point source 5C\,5.175 was observed throughout the observation in an interleaved manner and was used to correct for atmospheric and/or instrumental phase drift.
After narrow-band RFI flagging, the data were binned down to 64 channels to reduce processing time and imaged at the central frequency, 15.5\,GHz. We used the `clean' task, using multi-frequency synthesis with nterms=2 which allows for a frequency dependence of the sky brightness. We used the CASA graphical Gaussian fitting task on the resulting image to confirm that the source was unresolved and 
measure a peak flux density of  signal of $\textrm{S}_\textrm{$1.9\,$cm}\,=\,3.46\,\pm\,0.09\,\textrm{mJy}$ including thermal noise and a 5\% systematic error estimate at $15.5$GHz ($1.9$cm). \citealp{hurley-walker_bayesian_2012} (HW12) also observed cluster A851 in 2012 with AMI-LA reporting S$_\textrm{LA} = 2.2\pm 0.1$ at the position of {\dg } (Table 12, ID B in HW12) we include both results in Table~\ref{tab:observations}.

\begin{center}
\begin{table*}
\caption{Observed bands for Abell~851, the flux density (S) and uncertainty ($\delta$S) for each wavelength is provided in column 3 and 4.}
\label{tab:observations}
\begin{tabular}{lcccclc}
\hline
Instrument & Angular Resolution & Wavelength & S & $\delta$S & Reference & Observation Date\\
& & & (mJy) & (mJy) & & \\
\hline
WISE & 6.1$''$ & $3.4\micron$ & $\phantom{<}0.48$& $0.02$ &\citealp{wright_wide-field_2010} & 2010-2013 \\ 
WISE & 6.4$''$ & $4.6\micron$ & $\phantom{<}0.31$& $0.02$ & \citealp{wright_wide-field_2010}  & 2010-2013 \\
WISE & 6.5$''$ & $12\micron$ & $\phantom{<}0.60$& $0.17$ & \citealp{wright_wide-field_2010} & 2010-2013 \\
WISE & 12$''$ & $22\micron$ & $<3.0$& --- & \citealp{wright_wide-field_2010}  & 2010-2013 \\
PACS & 8$''$ & $100\micron$ & $\phantom{<}23.7$ & $0.3$ & \citealp{rawle_relation_2012} & 2003-07-01\\
PACS & 13$''$ & $160\micron$ & $\phantom{<}25.3$ & $0.7$ & \citealp{rawle_relation_2012} & 2003-07-01\\
SPIRE & 18$''$ & $250\micron$ & $\phantom{<}20.5$ & $5.5$ & \citealp{rawle_relation_2012} & 2003-07-01\\
SPIRE & 25$''$ &$350\micron$ & $\phantom{<}12.3$ & $6.2$ & \citealp{rawle_relation_2012} & 2003-07-01\\
SPIRE & 36$''$ & $500\micron$ & ${<}11.0$ & --- & \citealp{rawle_relation_2012} & 2003-07-01\\
SCUBA & 14$''$ & $850\micron$ & $<4.6$ & ---  & \citealp{smail_nature_2002} & 2000-11-27\\ 
IRAM PdBI & 6$''$  & $3.6\,$mm & $\phantom{<}2.59$ & $0.40$ & This work & 2014-06-07 \\ 
BIMA & 23$''$ & $1.05\,$cm & $\phantom{<}1.06$ & $0.15$  & \citealp{coble_radio_2007} & 1997-07-15\\ 
AMI-LA & 30$''$ & $1.9$cm & $\phantom{<}2.2$ & $0.1$ & \citealp{hurley-walker_bayesian_2012} & 2008-2009\\
AMI-LA & 30$''$ & $1.9$cm & $\phantom{<}3.46$ & $0.09$ & This work & 2017-10-04\\
VLA & 4$''$ & $6.2\,$cm & $\phantom{<}3.01$ & $0.14$ & Archive\protect\footnotemark[3] & 1990-12-01 \\
VLA & 4$''$ & $6.2\,$cm & $\phantom{<}2.79$ & $0.03$ & Archive\protect\footnotemark[3] & 1993-08-28 \\ 
VLA & 4$''$ & $21\,$cm & $\phantom{<}1.6$ & $0.16$ & \citealp{morrison_radio_1999} & 1996-01-06 \\ 
VLA/FIRST & 4$''$ & $21\,$cm & $\phantom{<}2.13$ & $0.31$ & \citealp{white_catalog_1997} & 1997-03-25\\
VLA & 4$''$ & $21\,$cm & $\phantom{<}3.1$ & $0.2$ & \citealp{condon_1.4_1990} & 1987-05-23 \\
\hline
\end{tabular}
\end{table*}
\end{center}

\section{Analysis and discussion} \label{sec:discussion}

\subsection{Spectral energy distribution}

\begin{figure}
\includegraphics[width=\columnwidth]{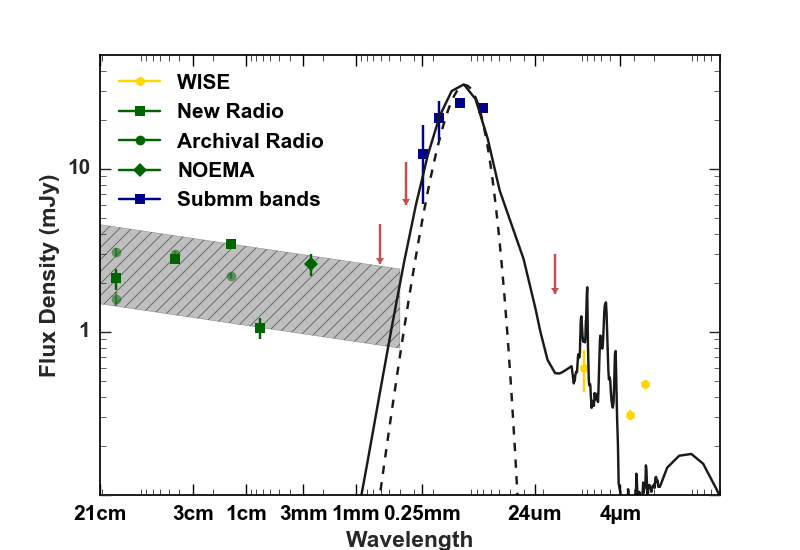}
\caption{The rest-frame spectral energy distribution for {\dg }. The blue squares cover the sub-mm wavebands: 100, 160, 250, 350, 500, 850$\micron$, the green diamond is the imaged IRAM PdBI 3.6mm detection, the gold hexagons are the FIR WISE bands, green squares are the most recent radio observations whereas the fainter green circles are historic observations (listed in Table~\ref{tab:observations}).
The shaded region shows the synchrotron emission amplitude range of a source with $\alpha\,\approx\,-0.1$ over two decades, the black dashed line is thermal blackbody component fit with a characteristic temperature of T$\,=\,24\,$K and, finally, the black solid line is a SPIRE $250\micron$ normalised template with $\alpha = 3.125$ from \citealp{dale_infrared_2002}}
\label{fig:sedFit}
\end{figure}

In Figure~\ref{fig:sedFit} we construct the spectral energy distribution using the data in Table~\ref{tab:observations} and fit three components spanning the radio, sub-mm and far-IR bands. Note that, despite the data spanning a range of angular resolutions, all observations are unresolved for \dg{} and a comparison of the beam sizes to the optical imaging of \dg{} shows that we are in all cases measuring galaxy-integrated flux densities with negligible contamination from neighbouring or background sources.

The \textit{Herschel} $100\,-\,350\micron$ flux measurements allow for a simple leaast squares fit of an isothermal modified blackbody, where we employ a standard emissivity term, $\beta=1.5$ \citep{hildebrand_determination_1983, casey_far-infrared_2012}, allowing dust temperature as the free parameter. We find a best-fit dust temperature of $T_D = 24$\,K, consistent with the S0 morphological--temperature results found by \citealp{bendo_dust_2003}. We also make use of the well known FIR/sub-mm templates described by \citealp{dale_infrared_2002} to fit the $100\,-\,350\micron$ data. We normalise to the $160\micron$ PACS detection as it lies near the peak of the thermal emission and we overlay the template which is best suited to the \textit{Herschel} data (corresponding to a power law index of $\alpha\,=\,1.875$), as a solid black curve. 

In the radio bands the spectral index of the synchrotron power law component is estimated from the ratio of the most recent NOEMA 3\,mm and the average AMI 1.9\,cm observations, giving $\alpha\,\approx\,-0.1$ for S\,$\propto\,\nu^{\alpha}$. We overlay a grey hashed region onto Figure~\ref{fig:sedFit} to represent the range of flux normalisations that fit the full range of radio observations up to the Rayleigh-Jeans tail of the thermal dust emission. While it is very likely that the synchrotron component will have some spectral curvature over the three orders of magnitude in frequency plotted \citealp{hogan_high_2015}, as we lack simultaneous observations over the full spectral range then we must interpret any variability assuming a single power law index, as we discuss in the next section.

\subsection{Variability}

\begin{figure}
\centering
\includegraphics[width=\columnwidth]{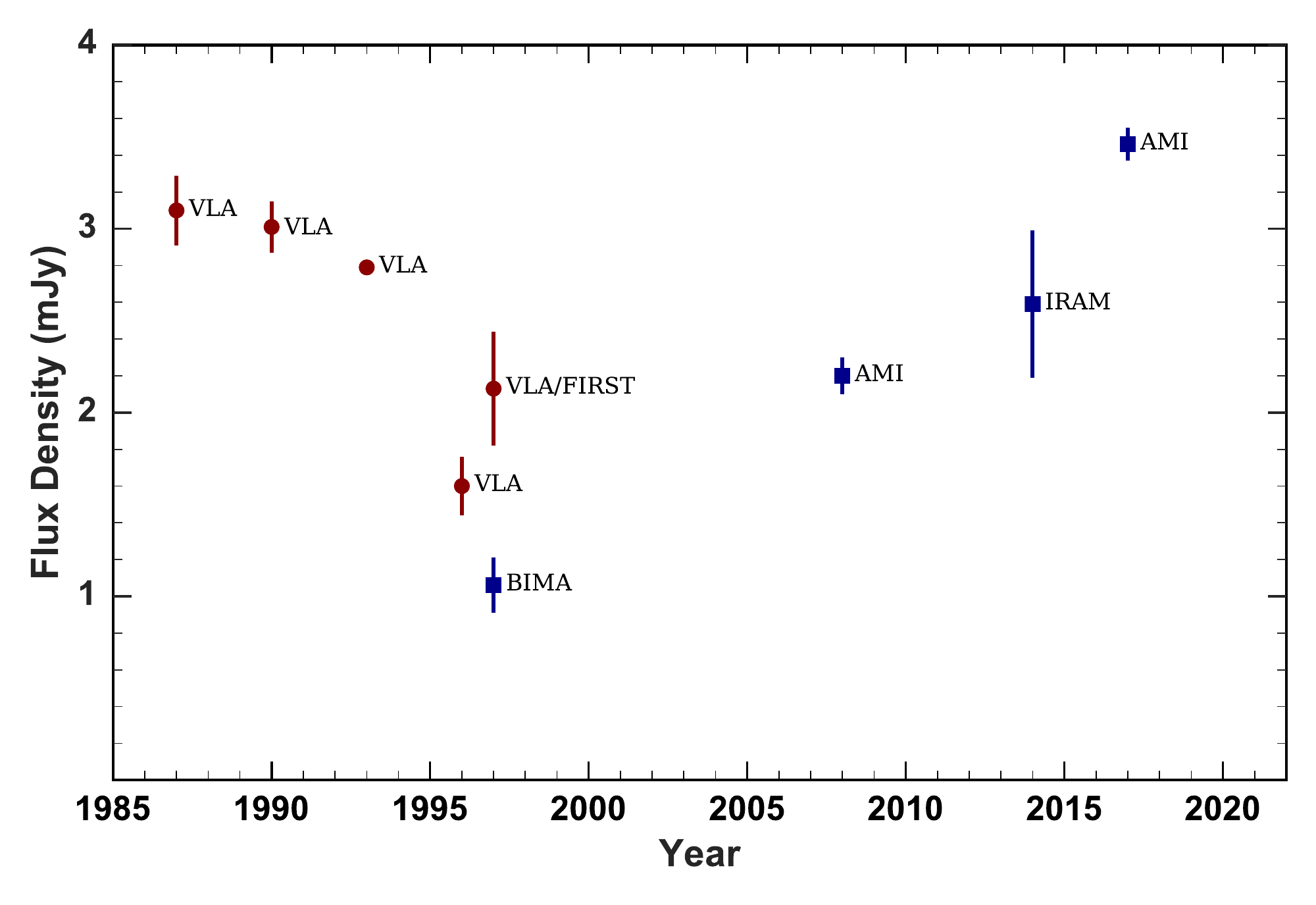}
\caption{Two decades of radio data from VLA, BIMA, AMI and IRAM reveals an up-turn in the BCG light curve. We derive out spectral index from the most recent observations by AMI and IRAM-PdBI. Red squares indicate data $\geq\,10\,$\,GHz and, blue circles $<\,10\,$\,GHz. It is clear from the figure that at $<\,10\,$\,GHz there has been a gradual decline in luminosity but in bands $\geq\,10\,$\,GHz there is evidence of an increase.}
\label{fig:vary}
\end{figure}

The radio light curves with observations from VLA, BIMA, AMI and NOEMA spanning near two decades in time is shown in Figure~\ref{fig:vary}. The source exhibits a flat spectrum over 1.4--82\,GHz range and is unlikely to be extended on the $>$few arcsecond scales of the resolution of the majority of observations (Table\ 1). Although none of these data were taken simultaneously, it is clear that over the 1.4--82\,GHz frequency range covered there has been a significant change in the normalisation of the power-law synchrotron component over the past 20--30 years. The source varies at the same frequency and resolution in data from both the VLA at 1.4\,GHz and AMI-LA at 15\,GHz; furthermore, the NOEMA data has better angular resolution than the BIMA and AMI-LA data but is brighter than all but the most recent AMI-LA point. If flux were being missed on larger scales, then one would expect the 3\,mm point to be fainter. The source appears to have been at a minimum around the year 2000 and has gradually increased in brightness by a factor of more than three in the past 18 years. We argue that the observed radio variability is intrinsic to the source.

This magnitude and timescale for radio variability is observed in local BCGs such as NGC1275 in the Perseus cluster \citep{dutson_non-thermal_2014} and a sample of bright, flat spectrum cores monitored with the OVRO 40m \citep{hogan_high_2015}. All of these variable BCGs are found in dusty BCGs with luminous optical lines in cool core clusters \citep{hogan_high_2015} with no evidence for jet contribution. As can be seen from the {\it HST} optical imaging in Figure~\ref{fig:pbCor}, \dg{} does exhibit dust lanes and the bright sub-mm detections across the {\it Herschel} bands indicate the presence of dust in the interstellar medium. \dg{} does not exhibit any significant H$\alpha$ line emission \citep{koyama_red_2011} and, although there is no {\it Chandra} observation of A851 and \dg{} falls on a chip-gap of the EPIC camera in the only on-axis pointing with {\it XMM-Newton}, the wider structure of the cluster in the X-ray from ROSAT \citep{schindler_x-ray_1998} suggests that any cool core in A851 is weak. On this basis \dg{} is atypical of BCGs with a strong radio core. The observed radio power of \dg{} of $\approx 10^{24}$~W~Hz$^{-1}$ implies that it falls in the upper quartile of core radio power of all X-ray luminous clusters \citep{hogan_comprehensive_2015} with a cool core or in the uppermost 3\% of core radio power for BCGs without a surrounding cool core. While the observed radio variability could be due to a jet, the similarity in its amplitude and time-scale implies a similar origin for the core emission in other BCGs. Therefore \dg{} is an important system that may have an unusual X-ray environment, and therefore more detailed X-ray follow-up is required to determine the properties of the intracluster gas on scales of 10s\,kpc around the BCG.

\section{Conclusions}

Optically, \dg{} appears to be a relatively unremarkable, dusty early-type disc galaxy (Sa/S0) but when observed in the radio--submillimetre this optically inactive galaxy appears to contain a relatively powerful, variable LLAGN in what is the Brightest Cluster Galaxy in a rich cluster. While the variability of a factor of three on decade timescales is consistent with other BCGs, the lack of a prominent cool core in the host cluster is surprising and highlights the need to assess the temporal behaviour of all massive galaxies in cluster cores, particularly in the millimetre wavelengths, to ascertain the underlying level radio mode regulatory feedback in massive clusters.

\section*{Acknowledgments}
RAC thanks Tim Pearson for useful comments and the anonymous referee for useful discussion. 
RAC is supported by the Royal Society, JEG is supported by a Royal Society University Research Fellowship. 
ACE acknowledges support from STFC grant ST/P00541/1.
YCP is supported by a Trinity College  JRF.
This work is based on observations carried out under project number S14BV with the IRAM PdB Interferometer. IRAM is supported by INSU/CNRS (France), MPG (Germany) and IGN (Spain). We thank the staff of the Mullard Radio Astronomy Observatory for their invaluable assistance in the operation of AMI, which is supported by Cambridge University and the STFC.
This research has made use of the NASA/IPAC Extragalactic Database (NED) which is operated by the Jet Propulsion Laboratory, California Institute of Technology, under contract with the National Aeronautics and Space Administration.



\bibliographystyle{mnras}
\bibliography{Zotero.bib} 


\bsp	
\label{lastpage}
\end{document}